\documentclass[aps,pre,twocolumn,groupedaddress,showpacs,floatfix]{revtex4}
\pdfoutput=1
\usepackage{graphicx}
\usepackage{amsmath}
\usepackage{amssymb}
\begin{document}
\title{Aftershocks in a frictional earthquake  model}
\author{O.M. Braun}
\affiliation{Institute of Physics, National Academy of Sciences of Ukraine,
  46 Science Avenue, 03028 Kiev, Ukraine}
\affiliation{International School for Advanced Studies (SISSA),
Via Bonomea 265, 34136 Trieste, Italy}
\author{Erio Tosatti}
\affiliation{International School for Advanced Studies (SISSA),
Via Bonomea 265, 34136 Trieste, Italy}
\affiliation{CNR-IOM Democritos National Simulation Center,
Via Bonomea 265, 34136 Trieste, Italy}
\affiliation{International Centre for Theoretical Physics (ICTP),
Strada Costiera 11, 34151 Trieste, Italy}

\begin{abstract}

Inspired by spring-block models, we elaborate
a ``minimal'' physical model of earthquakes
which reproduces two main empirical seismological laws,
the Gutenberg-Richter law and the Omori aftershock law.
Our new point is to demonstrate
that the simultaneous  incorporation of
ageing of contacts in the sliding interface
and of elasticity of the sliding plates constitute the minimal
ingredients to account for both laws within the same frictional model.
\end{abstract}
%
%
\maketitle

\section{Introduction}

Two very well known, empirically established laws of
planetary scale friction (i.e.,  seismology)
are the Gutenberg-Richter (GR) law~\cite{GR} and the Omori law~\cite{O1894}.
The former states that
the number of earthquakes (EQs) with magnitude $\geq {\cal M}$
scales with ${\cal M}$ as
\begin{equation}
{\cal N}({\cal M}) \propto 10^{-b{\cal M}} \propto {\cal A}^{-2b/3} ,
\label{eq01}
\end{equation}
where the dimensionless magnitude is defined as
\begin{equation}
{\cal M} = \frac{2}{3} \log_{10} \left( {\cal A}/{\cal A}_0 \right) ,
\label{calM}
\end{equation}
${\cal A}$ is a measure of the EQ amplitude (energy released, stress drop, etc.),
${\cal A}_0$ is a constant,
and $b \approx 1$ (or more generally $b = 0.5 - 1.5$~\cite{R2003,GW2010}).
The Omori law 
describes the rate of aftershocks (in excess of the background value)
at time $t$ after the main event,
\begin{equation}
n(t) = K_{\rm O} / (\tau_c + t)^p \,,
\label{Omori-eq}
\end{equation}
where
$K_{\rm O}$ depends exponentially on the magnitude $\cal M$ of the main shock,
$\log_{10} K_{\rm O} \propto \cal M$
\cite{YS1990},
$\tau_c$ has a typical average value of about 7~hours,
and $p \approx 1$ (or more generally $p = 0.7 - 1.5$~\cite{UOM1995}).
A similar behavior (with $\Delta t \to -\Delta t$)
was also reported for foreshocks~\cite{KK1978,JM1979}.
Both the GR law and the Omori law were established through a
statistical analysis of observed EQs.
Although widely addressed and discussed theoretically%
~\cite{N1990,RB1996,S1999,HN2002,MAGLPRV2003,HK2011,BK1967, 
CL1989,OFC1992,SGJ1993,G1994,RK1995,HZK2000,
J2010,STK2011,KHKBC2012,BP2013,Pelletier2000,J2013},
a generally accepted frictional model whose solution simultaneously accounts for both laws
seems to be still lacking.

We build on the time honored spring-block EQ model
dating back to Burridge and Knopoff (BK)~\cite{BK1967}, 
and subsequent work~\cite{CL1989,N1990,OFC1992,SGJ1993,
G1994,HZK2000,J2010,STK2011,BP2013},
where two rough sliding plates are coupled by a set of contacts
which deform when the plates move relative to one another.
The contacts are frictional,
behaving as elastic springs
as long as their stresses are below some threshold,
breaking to reattach in a less-stressed state
when the threshold is exceeded.
The EQ amplitude ${\cal A}$ is typically associated with
the number of broken contacts
at a global slip sliding event---the shock.
The BK-type models do predict a GR-like power-law behavior,
but typically for some particular sets of model parameters
(see a detailed analyzes of BK-type models in 
Refs.~\cite{Pelletier2000,KHKBC2012}),
and generally only for a restricted interval of magnitudes 
$\Delta {\cal M} \alt 2$%
~\cite{HZK2000,J2010,STK2011,BP2013},
unlike the much broader one observed in real EQs, $\Delta {\cal M} > 6$~\cite{BR1995}.
Beyond that partial failure, the existing EQ models
do not describe spatial-temporal correlations
between different EQs and thus fail altogether to explain the Omori law.
Aftershocks have been generally related to
relaxation~\cite{HZK2000,J2010}, but that aspect
is still in need of proper integration with others in a single model~\cite{HN2002,HK2011}.

We undertake that integration in the present work,
where we show that a simultaneous description of
both the GR law and the Omori law
can be obtained by models that incorporate two main ingredients:
the elasticity of the sliding plates and the ageing of contacts between the plates.

\section{The model}
\label{model}

\subsection{The sliding interface as a set of macrocontacts}
\label{model-macrocontact}

As typical in EQ-like models, we assume two 
plates, the top plate (the slider) and the bottom plate (the base),
coupled by a multiplicity of frictional micro-contacts.
If an individual micro-contact (asperity, bridge, solid island, etc.)
have a size $r_c$, then its elastic constant may be estimated as
$k_c \sim \rho c_t^2 r_c$, where $\rho$ is the mass density and
$c_t$ is the transverse sound velocity of the material which forms the asperity.

Elastic theory introduces a characteristic distance $\lambda_c$
known as the elastic correlation length,
below which the frictional interface
may be considered as rigid~\cite{CN1998,PT1999,BPST2012}.
Roughly it may be estimated as
$\lambda_c \sim a_c^2 E /k_c$,
where $a_c$ is an average distance between the micro-contacts
and $E$ is the slider Young modulus.
A set of $N_c = \lambda_c^2 /a_c^2$ micro-contacts within the
area $\lambda_c^2$ constitutes
an effective macro-contact~\cite{BPST2012,BP2012}
with the elastic constant
$k = N_c k_c \alt E \lambda_c$.
The macro-contact 
is characterized by a shear force $F_i (u_i)$,
where $u_i$ is the displacement of the point on the slider to which the
$i$th macro-contact is attached.
The function $F_i (u_i)$ may be calculated with
the master equation approach~\cite{BP2008,BP2010}
provided the statistical properties of micro-contacts are known.
Because of a strong (Coulomb-like) elastic interaction
between the micro-contacts at distances $r < \lambda_c$,
the effective distribution of threshold values for
frictional micro-contacts reduces 
to a narrow Gaussian distribution,
i.e., it is close to $\delta$-function~\cite{BPST2012}.
In this case $F_i$ linearly increases with $u_i$
until the macro-contact undergoes
the elastic instability~\cite{BT2009,BP2010,BT2011,BPST2012},
when (almost all) micro-contacts break and the macro-contact slides.

Thus, we assume that each macro-contact 
(simply called contact from now on)
is characterized by a shear force
$F_i = k u_i$
and by a threshold value $F_{si}$.
The contact stretches elastically so long as $|F_{i}| < F_{si}$,
but breaks and slides when the threshold is exceeded.
When a contact breaks, 
its shear force drops to $F_{i} \sim 0$, and evolution continues from there,
with a new freshly assigned value for its successive breaking threshold.

The macro-contacts  are
elastically coupled through the deformation of the slider's bulk.
The elastic energy stored between
two nearest neighboring (NN) contacts $i$ and $j$
in a non-uniformly deformed slider
may formally be written as $\frac{1}{2}K (u_i - u_j)^2$,
where $K$ is the slider rigidity defined below in Sec.~\ref{model-elastic}.
Within this multiscale theory of the frictional interface,
the sliding proceeds through creation and propagation
of self-healing cracks treated as solitary waves~\cite{BPST2012,BP2012}.

\subsection{Elastic model of the sliding plate}
\label{model-elastic}

\begin{figure} 
\includegraphics[clip, width=8cm]{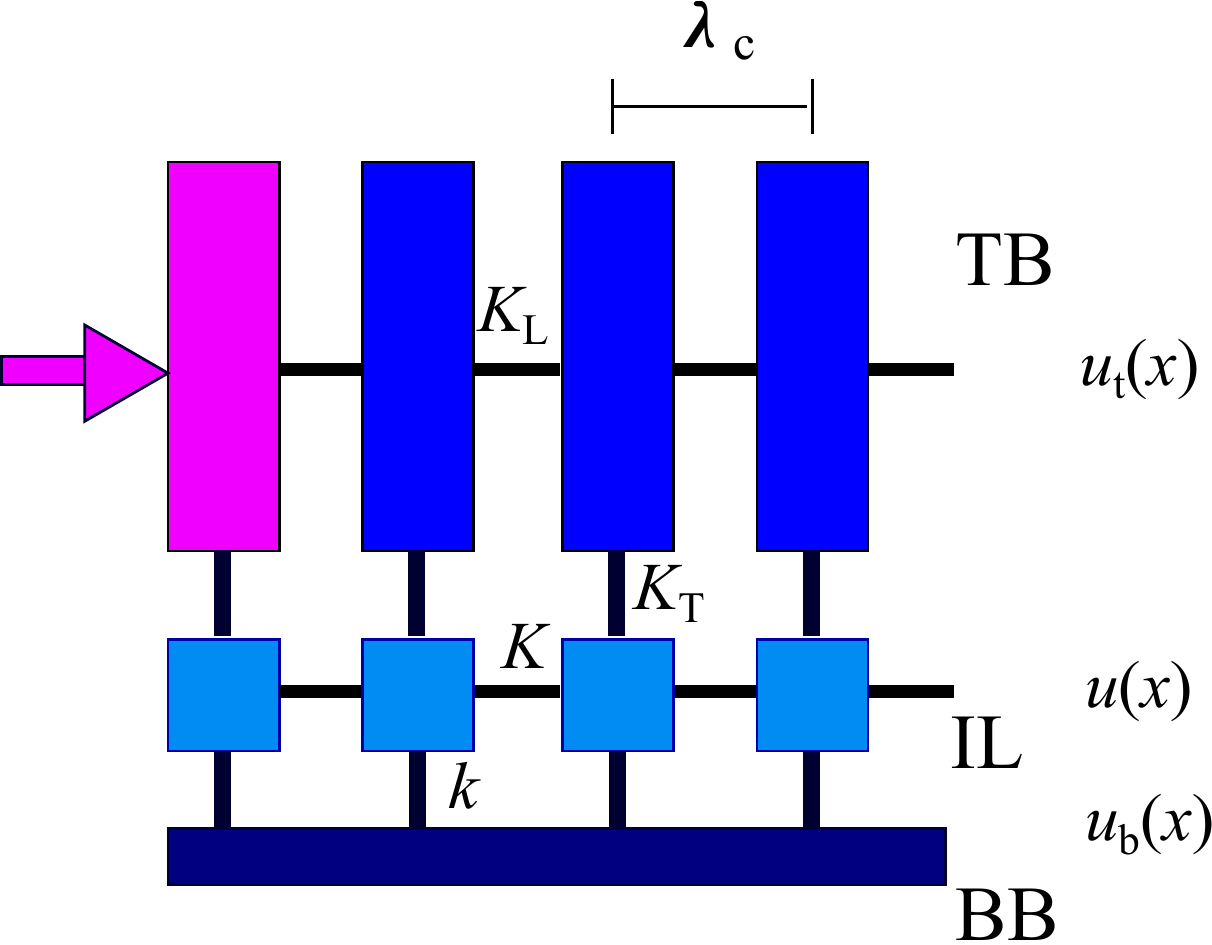} 
\caption{\label{A01}(color online): The model.
  The top block (TB) is split in rigid blocks of size
  $\lambda_c \times W \times H$
  connected by springs of elastic constant $K_{\rm L}$.
  The interface layer (IL) is split in rigid blocks of size
  $\lambda_c \times W \times \lambda_c$
  connected by springs of elastic constant $K$.
  The TB and IL are coupled by springs of elastic constant $K_{\rm T}$.
  The IL is connected with the rigid bottom block (BB)
  by contacts, represented by
  ``frictional'' springs of elastic constant $k$, which break when the local stress exceeds
  a threshold value.
}
\end{figure}

An earthquake corresponds to a release of elastic energy accumulated
in a body of the 
plate during its previous slow motion.
Therefore, any EQ model has to include the plate elasticity.
In a majority of EQ-like models this is done
indirectly through introducing an interaction between neighboring contacts.
For example, the most widely studied OFC model~\cite{OFC1992} assumes that
when the stress on one of the contacts reaches a threshold value $\sigma_s$,
it breaks, and the accumulated stress is equally redistributed over
the NN contacts, increasing their stresses on $\alpha \sigma_s$,
where $\alpha < 1/4$ is a parameter.
This may stimulate the NN contacts to break too, creating an avalanche
of contact breaking---a large shock.
The distribution of shock magnitudes in this case may follow the power law.
However, such explanation of the GR law is not ``robust''---%
the power law is observed for some sets of model parameters and 
for a restricted interval of EQ magnitudes,
typically much smaller than that observed in real EQs.
Instead, the GR law may be associated with macro-contact ageing alone
as discussed previously~\cite{BP2013} (see also Sec.~\ref{model-ageing}).

Nevertheless, the incorporation of the plate elasticity
into a realistic EQ model
is still important because, as we shall show in this work,
it is responsible for aftershocks.
When one of the macro-contacts breaks and slides,
the stress on neighboring contacts increases 
causing them to break as well,
and this process continues for some distance $\Lambda$,
where the contact stress drops
below than (but close to) its threshold value.
During propagation of this ``breaking wave''---%
corresponding to a large EQ, or the main shock---%
the previously accumulated elastic energy is released.
But an important issue is that the stress is not completely relaxed---%
the region where the wave was arrested, retains a
stress close to the threshold value
and thus represents a source for the next EQ---the aftershock.

The driving force which supports the propagation of the breaking wave,
is the stress accumulated in the body of the 
plate,
i.e., the latter plays the role of a ``stress reservoir''.
Because of the long-range character of stress distribution in an elastic body
(the stress decays with distance according to a power law),
this effect cannot be described by OFC-like models,
where the stress is localized;
therefore
one has to use a three-dimensional (3D) model of the plate.
A full fledge simulation of a 3D EQ model seems still out of
reach of modern computer power.
Instead, here we propose the two-layer model of the sliding plate, where
one layer (IL) plays the same role as in BK-type models,
while the second layer (TB) plays the role of a massive tectonic plate,
where the elastic stress is accumulated.

To be specific we elaborate here
the simplest one-dimensional (1D) version of the model,
where the contacts constitute a regular chain of length $L=Na$
with the spacing $a = \lambda_c$
as shown schematically in Fig.~\ref{A01}
(see Appendix~\ref{app-elastic} for reasons
how the 1D model may be ``deduced'' from the 3D elastic model of the plate
and the corresponding parameters,
Eqs.~(\ref{cra03}--\ref{cra05}) below, be defined).
The base (the bottom block BB) is assumed to be rigid,
whereas the elastic top plate
(the top block TB, the slider)
of length $L$,
width $W$ (in our 1D model we must set $W = \lambda_c$)
and an effective height (thickness) $H$
is modelled by the chain of
$N=L/a$ rigid blocks
coupled by springs of elastic constant
\begin{equation}
  K_{\rm L} = E \lambda_c \, (H/\lambda_c)
\label{cra03}
\end{equation}
(note that $H \gg \lambda_c$ must hold for
the correlation length $\lambda_c$ to be defined).
The contact between the TB and BB is described as an
interface layer (IL) of  thickness $H_c \sim \lambda_c$, consisting of $N$ contacts
coupled by springs with elastic constant
\begin{equation}
  K = E \lambda_c \,.
\label{cra04}
\end{equation}
The IL and TB are coupled elastically with
a transverse rigidity
which we model by $N$ springs of  elastic constant
\begin{equation}
  K_{\rm T} =
  \frac{E \lambda_c}{2(1+\sigma_{\rm P})} \frac{\lambda_c}{H} \;,
\label{cra05}
\end{equation}
where $\sigma_{\rm P}$ is the slider Poisson ratio.
Finally, the IL is coupled ``frictionally'' with the top surface of the BB---each
contact between the two plates is elastic with stiffness constant $k$
as long as the local shear stress in the IL is below the threshold.
The interface is stiff if $k \agt K$ and soft when $k \ll K$;
here we concentrate on the latter case.

Now, if a lateral pushing force is applied, for example,
to the left-hand side of the slider as shown in Fig.~\ref{A01},
the stress is transmitted to each contact due to the
elasticity of the slider.
In this case the interface dynamics starts by relaxation
of the leftmost contact which initiates the sliding.
This causes an extra stress on the neighboring contacts,
which tend to slide too, and
the domino of sliding events propagates as a solitary wave~\cite{BPST2012,BP2012},
extending the relaxed domain,
until the stress at some contact will fall below the threshold.
Such a self-healing crack propagates for some characteristic length $\Lambda$---%
leaving a relaxed stress behind its passage,
but raising the interface shear stress ahead of the crack.
The value of $\Lambda$ may be estimated analytically~\cite{BSP2014}; roughly
$\Lambda /a$ is proportional to $H/\lambda_c$ as well as to $k/K$.

\subsection{Ageing of the sliding interface}
\label{model-ageing}

The importance of incorporation of interface ageing---%
an effective strengthening of the interface due to slow relaxations,
growth of the contact sizes, or their gradual reconstruction,
chemical ``cementation'', etc.---%
is well known for EQ modelling as well as for tribological studies~\cite{KHKBC2012}.
In particular, ageing is held responsible for well-known effect---%
the transition from stick-slip to smooth sliding with changing of the
sliding velocity $v$~\cite{P0}.
Typically ageing is accounted for phenomenologically
by assuming that the friction force decreases when $v$ increases
(the velocity-weakening hypothesis).
At the simplest level,
one may just postulate the existence of two coefficients,
the static friction coefficient $\mu_s$ for the pinned state
and the kinetic friction coefficient $\mu_k < \mu_s$ for the sliding state.

In the EQ-like model, where the contacts continuously break and reform,
it is natural to assume that newborn contacts have initially
a small (e.g., zero) breaking threshold which then grows with
the timelife of the pinned contact.
In the simplest variant one may introduce simply a delay time
by assuming that the contact reappears after some time $\tau_d$
(so that for times  $t < \tau_d$ the threshold is zero).

In our model, where one macro-contact is represented
by many micro-contacts which break and reform,
it is natural to assume that the macro-contact
possess its own internal dynamics corresponding to a stochastic process.
Let $F_{si}$ be a threshold value for contact $i$.
Even assuming, as we will do, that newborn
contacts emerge with a vanishing  breaking threshold $F_{si} \sim 0$,
that threshold value will grow with time due to contact ageing.
We assume that the stochastic evolution of
contact thresholds is described by the simplest Langevin equation
\begin{equation}
dF_{si} (t)/dt = K(F_{si}) + G \, \xi (t) \,,
\label{eq31}
\end{equation}
where
$K (F_{si})$ and $G$ are the so-called drift and stochastic forces
respectively~\cite{Gardiner}, and
$\xi (t)$ is the Gaussian random force,
$\langle \xi (t) \rangle =0$ and
$\langle \xi (t) \, \xi(t') \rangle =\delta (t-t')$.
Equation~(\ref{eq31}) is equivalent to the Fokker-Planck
equation (FPE) for the distribution of thresholds ${\cal P}_c (F_{si}; t)$
(see Appendix \ref{app-ageing}):
\begin{equation}
\frac{\partial {\cal P}_c}{\partial t} +
\frac{dK}{dF_{si}} {\cal P}_c +
K \frac{\partial {\cal P}_c}{\partial F_{si}}
= \frac{1}{2} \, G^2 \frac{\partial^2 {\cal P}_c}{\partial F_{si}^2} \;.
\label{eq32}
\end{equation}

Our main assumption, Ref.~\cite{BP2013},
is that the drift force is given by
\begin{equation}
K(F_{si}) = \left( \frac{2\pi F_s}{\tau_0} \right)
\beta^2
\frac{1 - F_{si}/F_s}
{1 + \varepsilon (F_{si}/F_s)^2} \;,
\label{eq33}
\end{equation}
while the amplitude of the stochastic force is
\begin{equation}
G = (4\pi /\tau_0)^{1/2} \beta \delta F_s \;,
\label{eq34}
\end{equation}
where $\beta$ and $\varepsilon$ are two dimensionless parameters,
and the model parameters $\tau_0$, $F_s$ and $\delta \equiv \delta F_s/F_s$
define the space-time scale and are fixed below in
Sec.~\ref{simulation}.

With this choice, the stationary solution ${\cal P}_{c0} (F_{si})$ of
Eq.~(\ref{eq32})
corresponds to the Gaussian distribution
${\cal P}_{c0} (F_{si}) = (2\pi)^{-1/2}(\delta F_s)^{-1}
\exp \left[-\frac{1}{2} (1-F_{si}/F_s)^2/\delta^2 \right]$
in the case of $\varepsilon =0$,
while for $\varepsilon > 0$ the threshold distribution has a power-law tail,
${\cal P}_{c0} (F_{si}) \propto F_{si}^{-1/\varepsilon \delta^2}$ for $F_{si} \gg F_s$.
Therefore, the dimensionless parameter $\varepsilon$ determines the deviation
of the threshold distribution from the Gaussian shape,
while $\beta$ corresponds to the rate of ageing.

The stochastic dynamics alone leads to the power-law distribution of thresholds
in the stationary state.
Thus, if the plate moves adiabatically, $v \to 0$, then at $t \to \infty$
the distribution of EQ magnitudes will follow the GR-like law
even for ${\cal M} \to \infty$.
However, when $v > 0$, there is a competition between the growing process
and the continuous contact breaking due to sliding,
and the maximal magnitude of EQs is restricted.
For the steady state motion the corresponding solution may be found
analytically with the master equation approach~\cite{BP2013};
it gives
\begin{equation}
\label{eq:mmax}
{\cal M}_{\rm max} \propto \log_{10} \left( \beta /
  \sqrt{v} \right) \;.
\end{equation}

\section{Simulation}
\label{simulation}

In simulations we typically used a chain of $N = 901$
macro-contacts with periodic boundary condition.
Four parameters of our model may be fixed without loss of generality:
$a = 1$ (the length unit),
$K = 1$,
$m = 1$
(mass of the macro-contact),
and $F_s = 1$.
Then, the characteristic frequency is
$\omega_0 = (K/m)^{1/2} =1$, so that the unit of time is
$\tau_0 = 2\pi /\omega_0 =2\pi$.

\begin{figure} 
\includegraphics[clip, width=8cm]{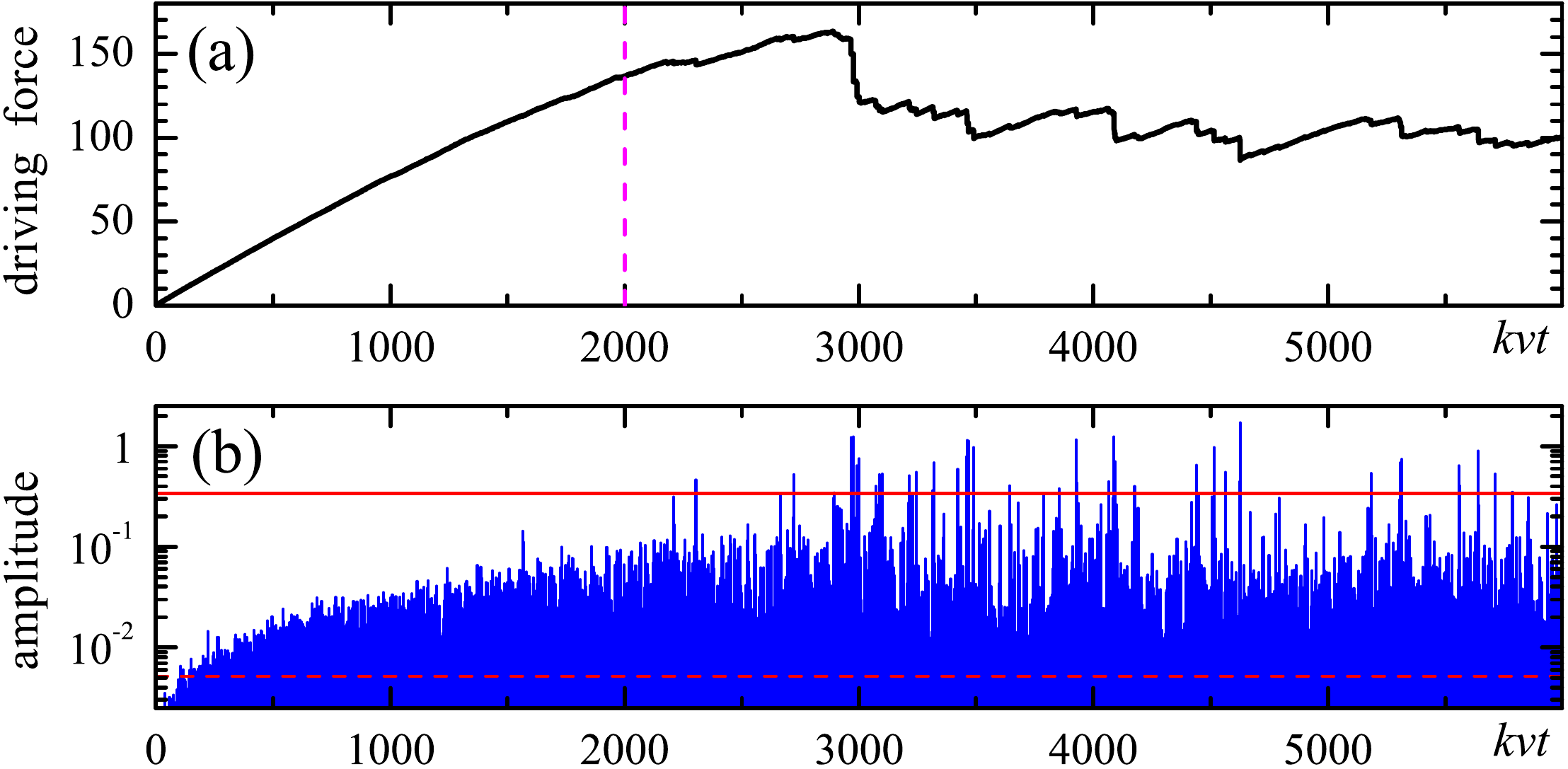} 
\includegraphics[clip, width=8cm]{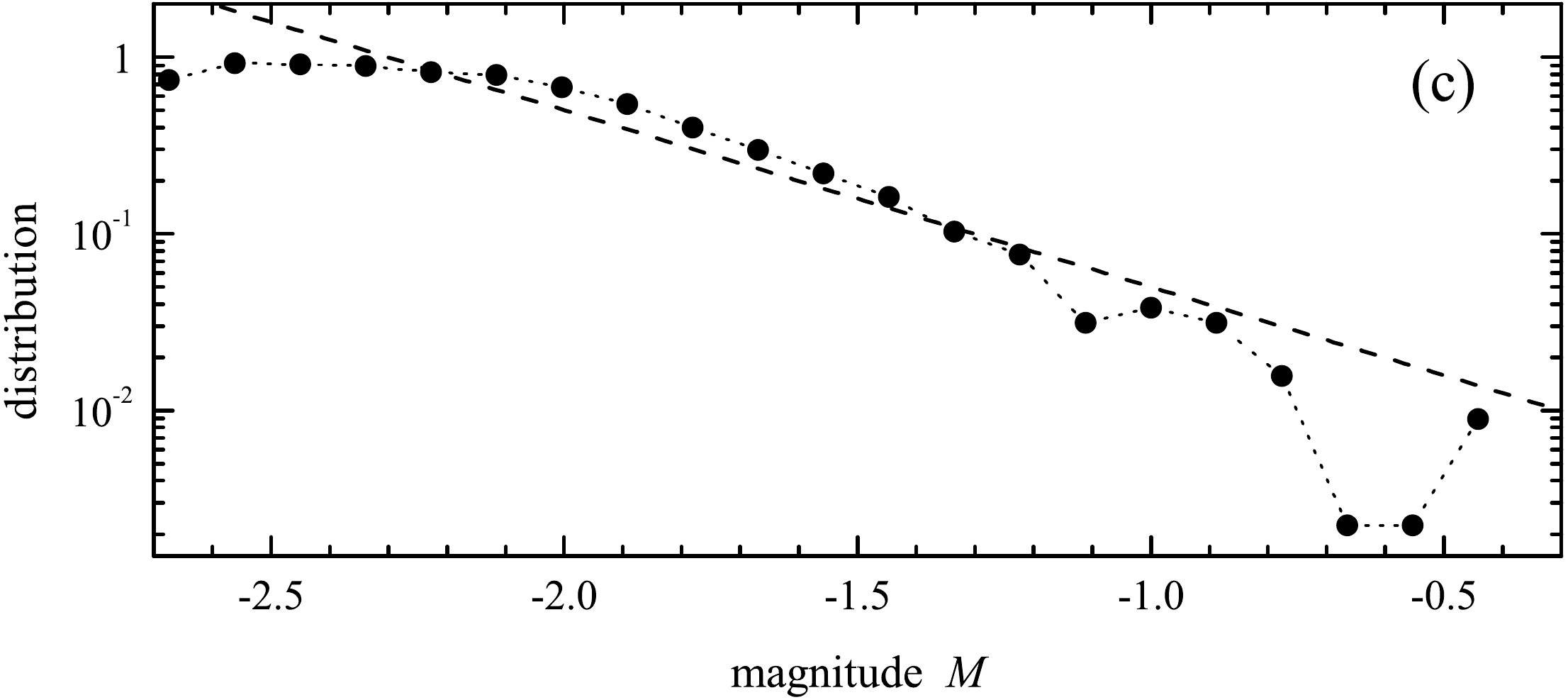} 
\caption{\label{A03}(color online):
Time evolution of the system:
(a)~the frictional 
force $F(t)$, and
(b)~the (global) amplitude of earthquakes ${\cal A} (t)$ versus time.
(c)~Statistics of earthquake magnitudes presented in~(b)
(the first 33\% of data discarded)
showing the Gutenberg-Richter power law behavior;
the dashed line corresponds to the exponent $b=1$.
}
\end{figure}

\begin{figure} 
\includegraphics[clip, width=8cm]{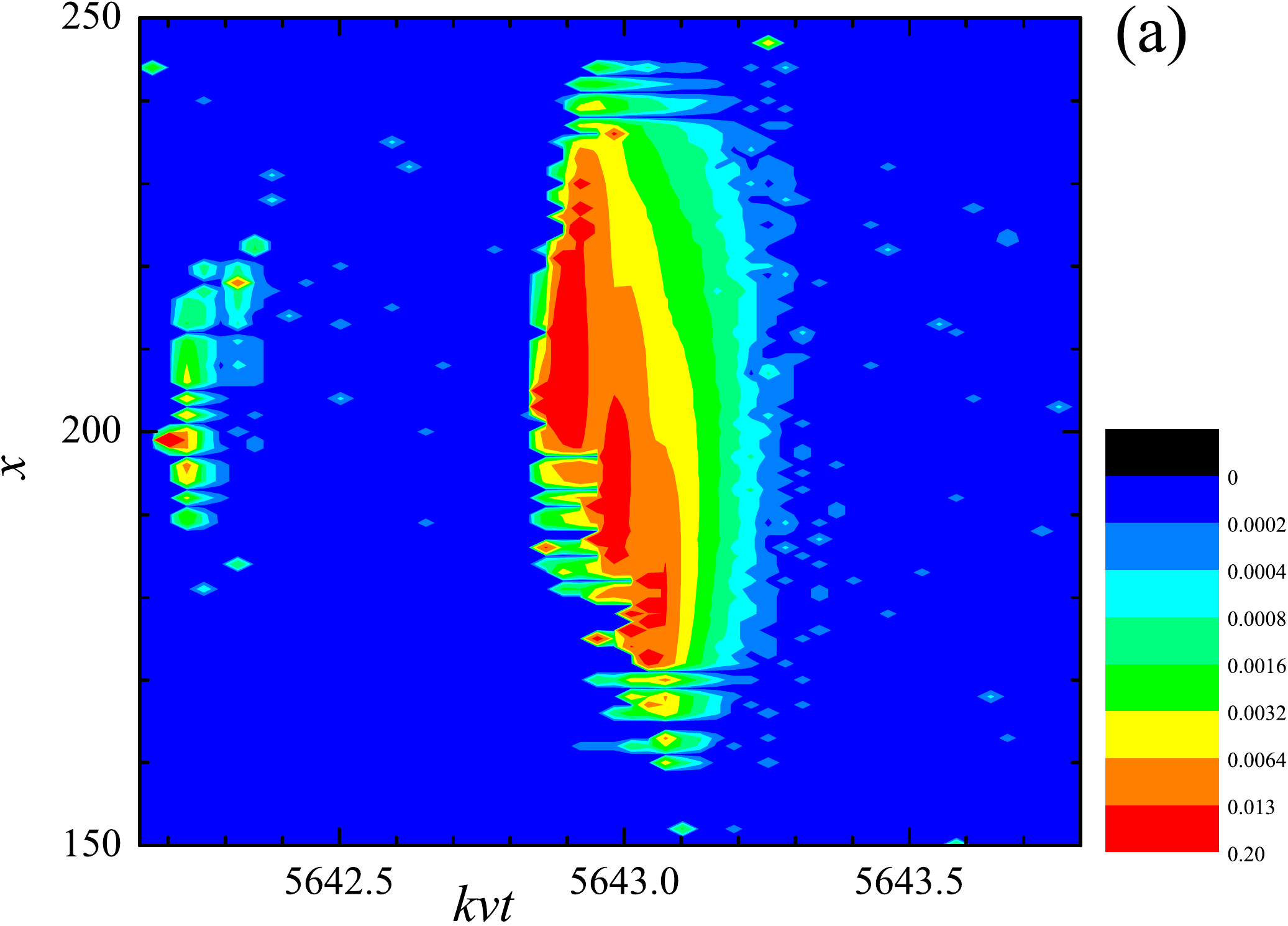} 
\includegraphics[clip, width=8cm]{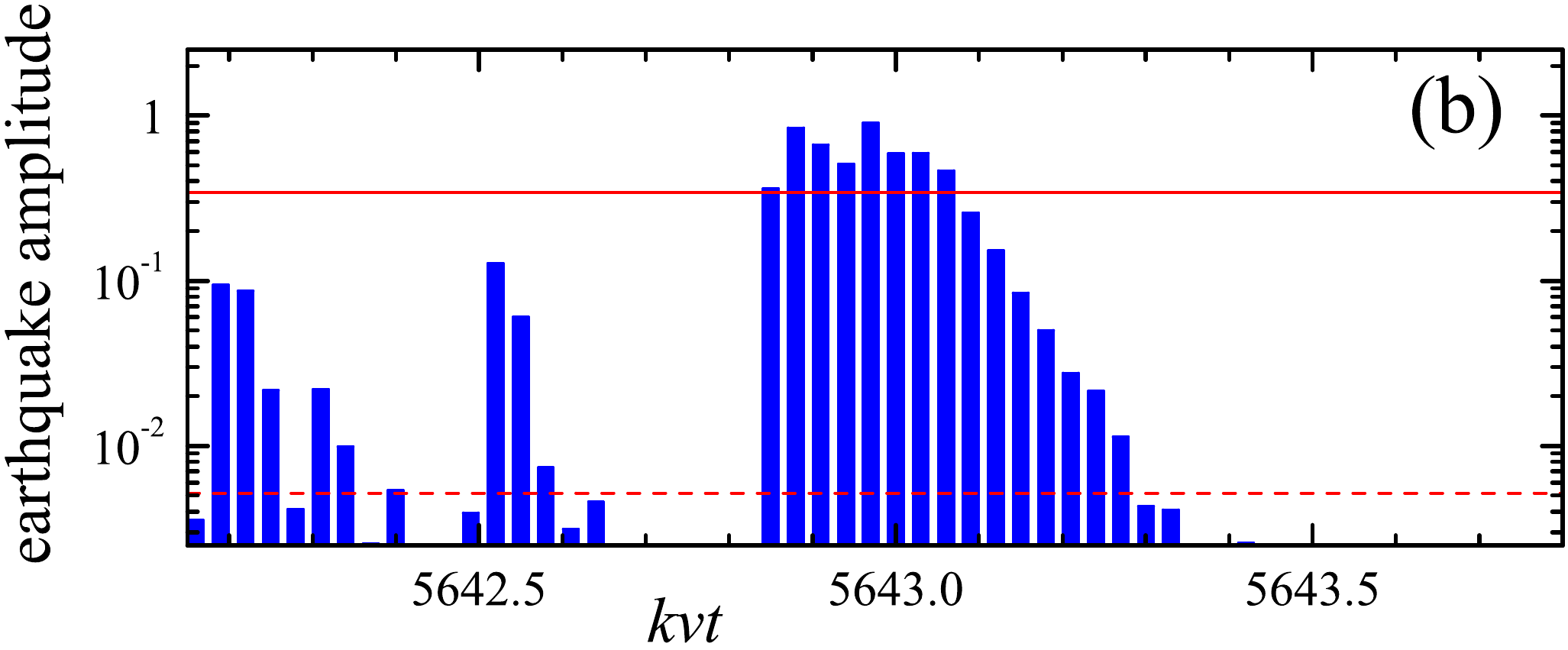} 
\caption{\label{A04}(color online):
Typical earthquake in our model:
(a)~the color map of the earthquake amplitude on the $(t,x)$ plane, and
(b)~the earthquake amplitude ${\cal A} (t)$ versus time.
}
\end{figure}

\begin{figure} 
\includegraphics[clip, width=8cm]{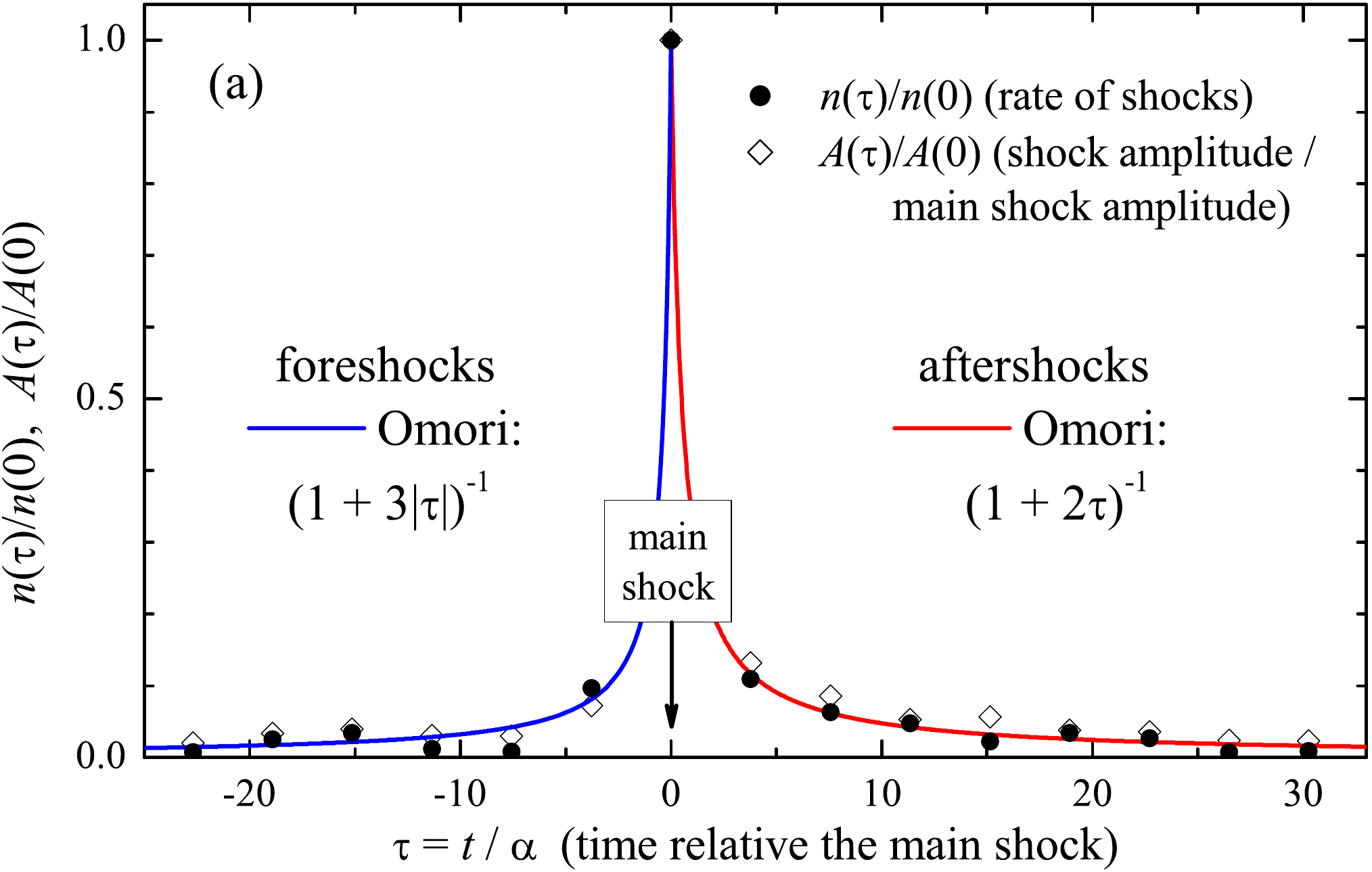} 
\includegraphics[clip, width=8cm]{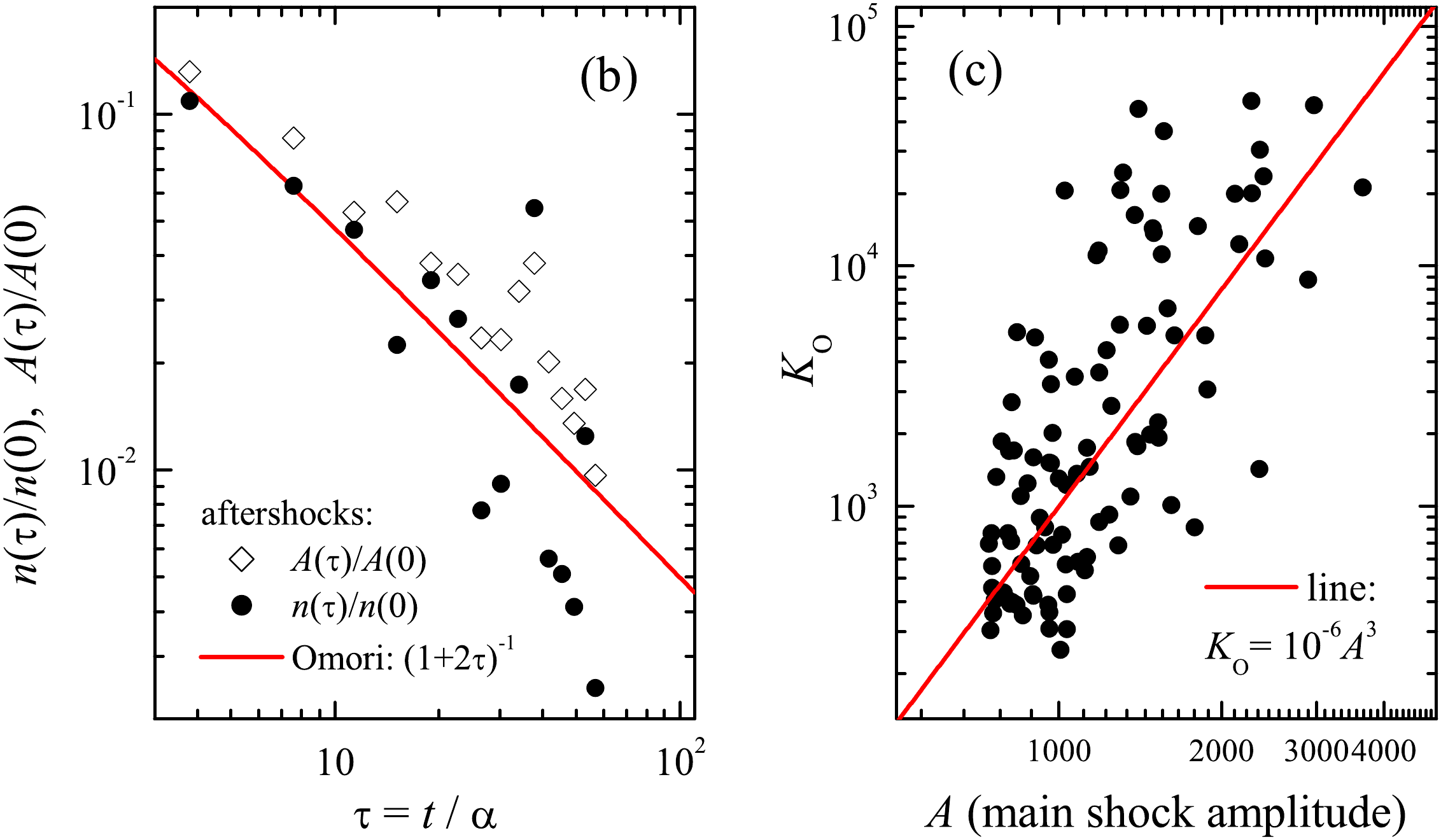} 
\caption{\label{A05}(color online):
(a) Foreshocks and aftershocks statistics:
the rate of fore- and aftershocks $n(\tau)$
(filled circles)
and the shock amplitudes ${\cal A} (\tau)$ (open diamonds)
relative the corresponding main shock.
The curves demonstrate the Omori law.
(b) Aftershocks statistics in log-log scale.
(c) The coefficient $K_{\rm O}$ in Eq.~(\ref{Omori-eq})
as function of the amplitude of the main shock in log-log scale.
The line shows the $K_{\rm O} \propto {\cal A}^3 \propto 10^{4.5{\cal M}}$ dependence.}
\end{figure}

For the elastic slider we used 
      $N_{\rm L} \equiv H/\lambda_c   = 100$ so that
      $K_{\rm L} = N_{\rm L} K$ and
      $K_{\rm T} = K/(2.6 \, N_{\rm L})$.
We took $\sigma_{\rm P} = 0.3$ for the Poisson ratio,
and $M = N_{\rm L} m$ for the mass of TB blocks.
The top layer of the slider is driven
through springs attached to the TB blocks, each with the elastic constant
      $K_d  = 0.03 \, K$,
the springs ends moving with the constant velocity
      $v = 0.01$. 
In simulations, the contacts elastic constant is
      $k    = 0.03 \, K$
and the thresholds dispersion parameter
$\delta F_s = 0.1 F_s$;
that yields a characteristic distance
for self-healing crack propagation
$\Lambda \alt 100 \, a$~\cite{BSP2014}.
A contact ageing parameter
$\varepsilon = 75$
yields a GR distribution with $b \approx 1$
(see Ref.~\cite{BP2013});
a reasonable rate of ageing is chosen as $\beta = 1$.
The newborn contacts emerge with a threshold $F_{si}$
taken from the Gaussian distribution with average
$F_{s, {\rm nb}} = 0.01 F_s$
and deviation $\delta F_{s, {\rm nb}} = F_{s, {\rm nb}}$.
In the initial state, all contacts are relaxed, $F_i =0$,
and all thresholds correspond to newborn contacts.
Then, the equations of motion for all blocks
(with a viscous damping coefficient $\eta = 1$)
together with the Langevin equations for contact breaking thresholds are solved
with a time step $\Delta t = 0.01$.
The total simulation time required is $t_{\rm max} = 2 \times 10^7$
(taking a typical velocity of a
tectonic plate $v \sim 30$~mm/year~$\approx 3 \times 10^{-9}$~m/s
and a distance between the contacts $a \sim 1$~mm,
we obtain that the time step $\Delta t = 0.01$
in simulation with $a=1$ and $v=0.01$
corresponds to a real-time step $\sim 30$~sec
and the total time $\sim 6 \times 10^{10}$~sec~$\sim 2 \times 10^3$~years).
In contrast to the cellular automaton algorithm
typically used in simulations of the Burridge-Knopoff model~\cite{BK1967},
in the present work we use, following Carlson and Langer~\cite{CL1989},
an alternative
algorithm with a fixed time step $\Delta t$.
Although in this case 
more than one contact breaking event may take place in a single time step
(which corresponds to an earthquake in our model with ``living'' macro-contacts)
that does not alter our main results
(e.g., see discussion in Ref.~\cite{BP2013}).
We define the (dimensionless) quake amplitude
as the sum of force drops on contacts at every time step,
\begin{equation}
{\cal A} (t) = \sum_i \Delta F_i (t)/F_s \,,
\label{calA}
\end{equation}
and the rate of shocks $n(t)$ as the number of shocks per one time step
(regardless of their amplitude).

The results are presented in Figs.~\ref{A03} and~\ref{A04}.
A typical time sequence of EQ shock sizes is that of Fig.~\ref{A03}b.
On large time scale, the function ${\cal A} (t)$ 
appears stochastic.
On a finer scale though (Fig.~\ref{A04}b)
one can distinguish separated shock blocks,
each with a large main shock followed by aftershocks.

\section{Analysis of results}

We can now analyze
the EQ shocks for their characteristics, distribution
and correlations in space and time.
To do that efficiently, we first remove from the analysis the small ``background'' EQs
with amplitudes below some level---we retain only
${\cal A} > 2 \, \langle {\cal A} (t) \rangle$
(broken red line in Figs.~\ref{A03}b and~\ref{A04}b).
Next, we single out the main EQs (MEQs) above some level
${\cal A}_{\rm MEQ}$; we took here
${\cal A}_{\rm MEQ} = 0.2 \, {\cal A}_{\rm max}$
(red line in Figs.~\ref{A03}b and~\ref{A04}b).
We also used the following rescaling procedure to distinguish from one another
MEQs that may occur too close:
If $n_{\rm MEQ}$ is the total number of MEQs,
call $\sigma = S / n_{\rm MEQ}$ the average
area occupied by a single MEQ on the $(t, x)$ plane, with $S = t_{\rm max} L$.
We then rescale the time coordinate
$t \to \tau = t/\alpha$ with $\alpha = \sigma /\lambda^2$, where $\lambda$
is some distance chosen in such a way that
the distribution of MEQs on the $(\tau, x)$ plane becomes {\it isotropic}
($\lambda \approx 235$ for the parameters used in Fig.~\ref{A03}).
We can now scan all MEQ coordinates on the $(\tau, x)$ plane and,
if the distance $\rho_{ij} =
[ (\tau_i - \tau_j)^2 + (x_i - x_j)^2 ]^{1/2}$
between two MEQs $i$ and $j$
is smaller than some value $\rho_{\rm cut}$
(we chose $\rho_{\rm cut} = 0.75 \, \lambda$),
then only the larger of these two MEQs
remains as the MEQ, while the lower one is removed from the list of MEQs.

With this protocol we
have obtained a set of well separated MEQs isotropically occupying the $(\tau, x)$ plane,
and we may calculate the temporal and spatial distribution of all earthquakes
within some area around every MEQ---we count the EQs separated
from the corresponding MEQ by less
than $\rho_0 = \rho_{\rm cut} /3$.
Then we collapse all data
together, designating $\tau =0$ for every main shock
and normalizing shocks amplitudes on the corresponding main shock value
(because the data so obtained are still noisy,
we coarsened the distribution with an extra width
$\Delta \rho$ in  the $(\tau, x)$ plane,
using $\Delta \rho = \rho_0 /31$).

The EQ distribution---our main result---is presented in Fig.~\ref{A05}.
One can clearly see that the aftershocks satisfy the Omori law
(Fig.~\ref{A05}).
The number of aftershocks [the coefficient $K_{\rm O}$ in the Omori law~(\ref{Omori-eq})]
depends exponentially on the magnitude ${\cal M}$ of the corresponding main shock
(Fig.~\ref{A05}c),
although the numerical coefficient of the exponent ($\sim 9/2$) is
essentially larger than that reported for real earthquakes
($\sim 1/2 - 2/3$, see~\cite{YS1990});
this may be connected with that the ``EQ amplitude'' ${\cal A}$
defined in our 1D model,
is not exactly equivalent to the seismic moment $M_{\rm O}$
used in analyzing of EQ statistics.
The spatial correlation between the shocks 
decays exponentially in our 1D model
unlike real earthquakes,
where the aftershock amplitude decays power-law with distance
owing to 3D elasticity~\cite{FB2006};
since the origin of that discrepancy is so very clear,
it does not worry us.
Foreshocks are also observed in simulation,
typically with a smaller value of the lag time $\tau_c$.

\section{Conclusions}

The spring-block frictional model elaborated and solved in this paper
confirms and details the mechanism through which elasticity of the sliding plates
and contact ageing enter as the
key ingredients of a minimal EQ model.
Without either of
them, a coexistence of the GR power law energy distribution with Omori's power
law aftershock distribution would not occur. In the
earlier model of Ref.~\cite{BP2013}
it was shown instead how that the GR earthquake law could arise due to contact ageing alone.
Explicit incorporation of the slider's elasticity done here provides also
the spatial-temporal distribution of EQs in fair accordance with the Omori aftershock law.

The magnitude of the largest EQ is controlled in the present model by the parameter $\beta$,
the rate of ageing relative the driving velocity.  The spatial radius of aftershock activity
is controlled by the length $\Lambda$, the self-healing crack propagation distance.

As a sobering remark in closing, stimulated by one Reviewer,
we should observe that the present modeling has no pretense
to represent the full complexity of real earthquakes.
In experimental aftershock sequences
the first aftershock 
has usually a magnitude one unity smaller
than the main shock (Bath's law~\cite{Bath1965})
whereas in the numerical result (Fig.~\ref{A04}b)
we have clusters of consecutive  earthquakes
with about the same amplitudes. 
Also, 
the dependence of the coefficient $K_O$ in the Omori law
on the main shock magnitude
is poorly reproduced in the present 1D model.
Besides, 
the true number of foreshocks is typically
much smaller than that obtained here (Fig.~\ref{A05}).

Future improvements of the model will clearly be needed to account
for these additional aspects, besides
the very basic ones which it presently deals with.
The study of a 2D model of the interface and
a 3D model of the elastic slider would provide a power-law EQ spatial distribution.
More generally
we think one can build on this basic modeling and to describe, through extensions
and adjustment of model parameters and ingredients, the greater complexity of real EQs.

\acknowledgments

This work is supported in part by COST Action MP1303.
O.B.~was partially supported by
PHC Dnipro/Egide Grant No.~28225UH
and the NASU program ``RESURS''.
Work in Trieste was sponsored through ERC Advanced Grant  320796  MODPHYSFRICT,
and by SNF Sinergia Contract CRSII2 136287.


\appendix
\section{Two-layer model of the elastic plate}
\label{app-elastic}

\begin{figure} 
\includegraphics[clip, width=8cm]{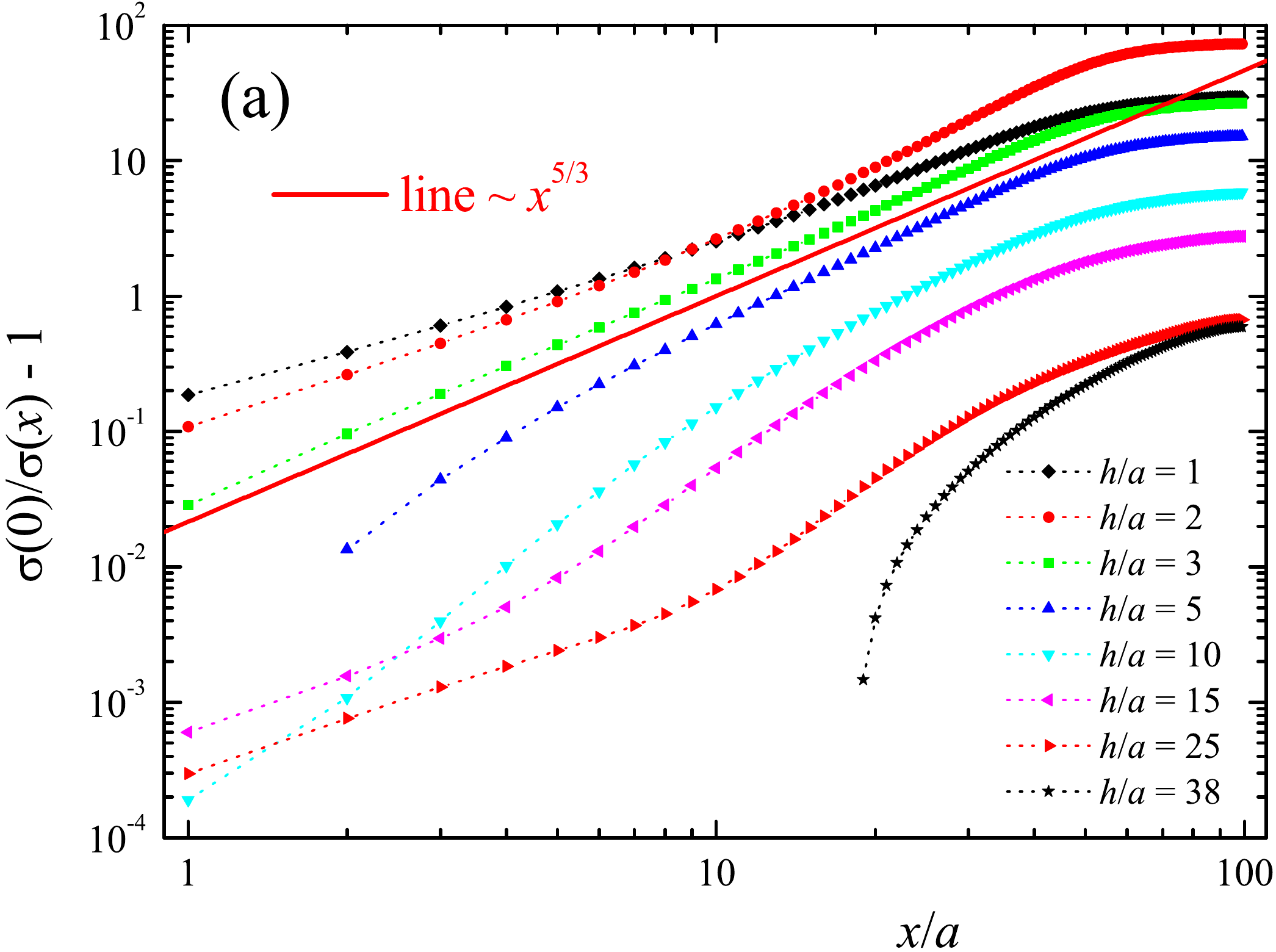} 
\includegraphics[clip, width=8cm]{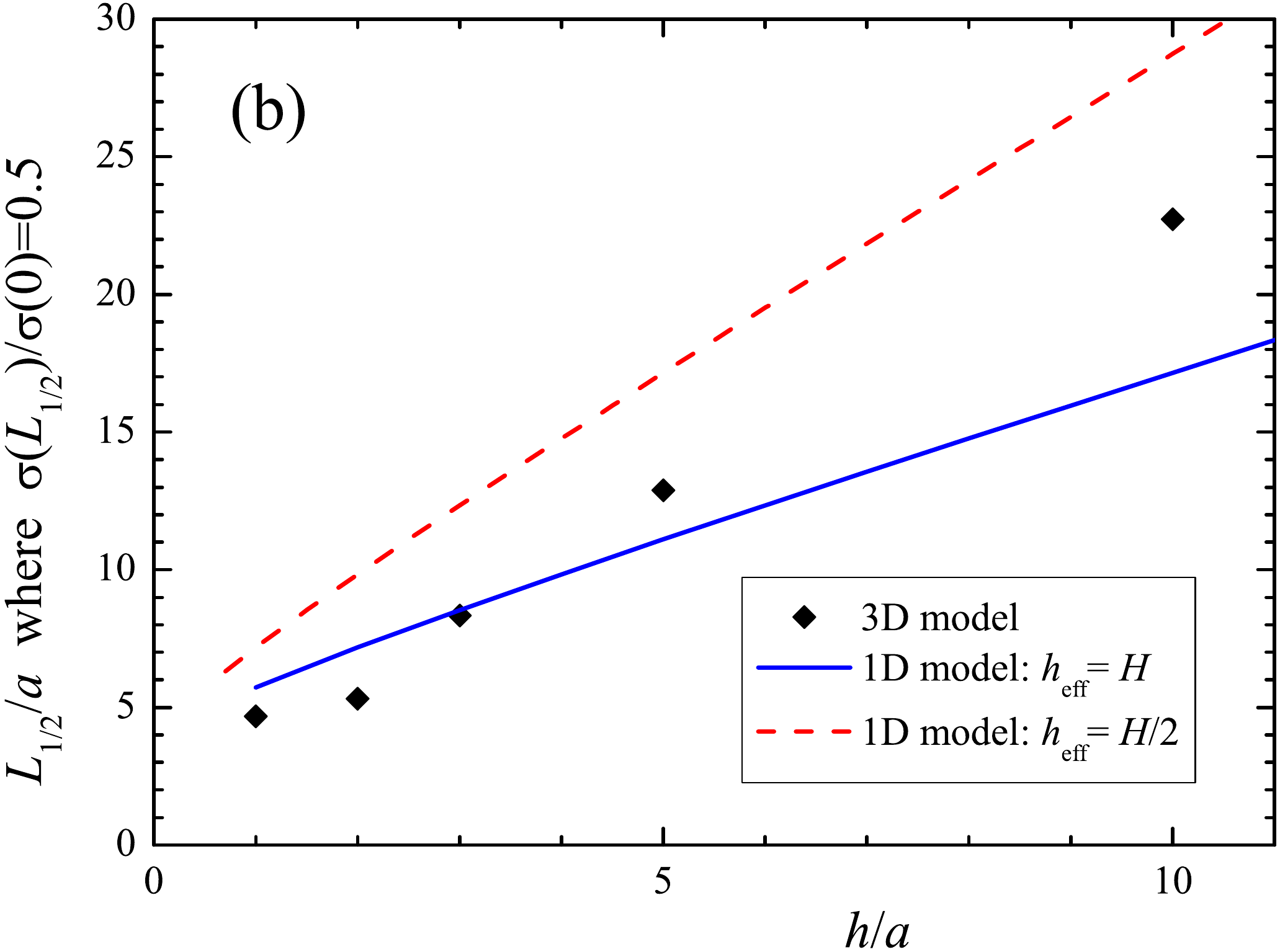} 
\caption{\label{A06}(color online):
(a) The shear stress $\sigma$ in the interface versus $x$
for the 3D slider of size $100a \times 3a \times 38a$ with
the mass density $\rho = 1.19 \times 10^3$~kg/m$^3$
and the sound speeds $c_l = 2680$~m/s and $c_t = 1100$~m/s
(the parameters corresponding to plexiglass
experimentally studied in Ref.~\cite{RBRBUF2011}),
$k/K=0.03$.
(b) Dependence of the length $L_{1/2}$ on the height of
the applied pushing force $h$ (symbols, numerics);
lines show the 1D model dependences $L_{1/2} (h_{\rm eff})$
with $h_{\rm eff} = H$ (blue solid)
and $h_{\rm eff} = H/2$ (red dashed curve).
}
\end{figure}

There are no way to reduce the 3D elastic model of the tectonic plate
to a 1D model rigorously, but one may define a 1D model
that leads to qualitatively reasonable results.
Let us consider the 3D elastic slider of size $L \times W \times \widetilde{H}$
with the longitudinal rigidity ${\cal K} = EW\widetilde{H}/L$
and the transverse rigidity
${\cal K}_{\rm T} = [E/2(1+\sigma_{\rm P})] (LW/\widetilde{H})$.
For numerical study, we split it into cubes of linear size $a$.
The interaction between the NN cubes
in the isotropic case is described by two constants,
the longitudinal elastic constant
$\kappa_l = \rho c_l^2 a$
and the transverse one,
$\kappa_t = \rho c_t^2 a$,
where $\rho$ is mass density and
$c_{l,t}$ is the longitudinal (transverse) sound speed~\cite{LLbook}.

Then, let the slider be coupled with the rigid bottom plate
by $LW/a^2$ springs of elastic constant $k$.
If we now apply the pushing force
from the left-hand side of the slider
at a hight $h < \widetilde{H}$, 
it will produce the stress $\sigma(x)$ at the slider/BB interface
(we consider the system uniform in the $y$-direction,
along which we apply the periodic boundary condition).
The function $\sigma(x)$ for the 3D slider roughly follows a power law,
$\sigma(x) \propto x^{-\nu}$,
where the exponent $\nu <3$ depends on model parameters
(see Fig.~\ref{A06}a).
To characterize the decaying function $\sigma(x)$,
we define the length $L_{1/2}$ where the stress decreases
in two times, $\sigma(L_{1/2})/\sigma(0) =0.5$.
Figure~\ref{A06}b shows the dependence of $L_{1/2}$ on the height $h$
where the pushing force is applied.
Of course, a 1D model cannot reproduce the power-law dependence $\sigma(x)$,
the 1D model always gives the exponentially decaying stress distribution.
Our idea is to construct an effective 1D model that
leads to a qualitatively similar dependence of
$L_{1/2}$ on model parameters.

Let us consider the lowest layer of the slider
as the IL so that it corresponds to a chain of $N=L/a$ cubes
coupled with the BB by springs $k$ and connected between themselves
by springs $K=\kappa_l =Ea$, Eq.~(\ref{cra04})
(in a general case one may put $K=\kappa_l E_c/E$,
where $E_c$ is the Young modulus of the interface,
but we do not see reasons to introduce additional parameters
in our qualitative 1D model).

Then, let us connect rigidly in the $z$-direction
the remaining $N_z =H/a$ ($H = \widetilde{H} -a$) layers of the slider,
so that it now consists of $N$ rigid blocks
coupled by springs of elastic constant $K_{\rm L} = N{\cal K} = EH$, Eq.~(\ref{cra03}),
as shown in Fig.~\ref{A01}.
Finally, to reproduce the slider transverse rigidity ${\cal K}_{\rm T}$,
we couple the IL and TB by $N$ springs of  elastic constant
$K_{\rm T} = {{\cal K}_{\rm T}}/{N}$ which gives Eq.~(\ref{cra05})
(one may think that we have to put $K_{\rm T} =\kappa_t$,
but this choice does not simulate correctly the slider transverse rigidity
and does not reproduce the dependence of $L_{1/2}$ on model parameters
qualitative similar to that of 3D numerics).

Thus, we came to the effective 1D model of the slider
described in Sec.~\ref{model-elastic}, Fig.~\ref{A01}.
This model allows us to find analytically the stress distribution along the interface,
$\sigma(x) \propto \exp (-\kappa_2 x)$, where
\begin{align}
  & \kappa_{2}^2 =
  \frac{1}{2} \left[ (\kappa^2 + \kappa_T^2) - \sqrt{\cal D} \, \right] ,
  \label{cra19}
\\
  & {\cal D} = \left( \kappa^2 - \kappa_T^2 \right)^2 + 4 \beta \kappa^2 \kappa_T^2 \,,
  \label{cra21}
\\
  & (a \kappa_T )^2 = K_T /K_L \,,
  \label{cra11}
\\
  & (a \kappa)^2 = (k + K_T)/K \,,
  \label{cra12}
\\
  & \beta = K_T / (K_T + k \, ) \,,
  \label{cra13}
\end{align}
so that now $L_{1/2} = (\ln 2)/\kappa_2$.
The comparison of the 1D analytical dependence $L_{1/2} (H)$
with the 3D numerical one $L_{1/2} (h)$ is presented in Fig.~\ref{A06}b.
One can see that, if we interpret the parameter $H$ in our 1D model
as some effective thickness of the tectonic plate
where the elastic stress is accumulated,
we could obtain qualitative correct results.
Moreover, the 1D model allows us to find analytically
the characteristic length $\Lambda$---%
the propagation path of self-healing crack---%
which determines the distance where the stress remains unrelaxed
(and even increases to become close to the threshold)
after a large shock and, therefore, where an aftershock could occur.

Thus, our 1D model of the slider is characterized by two dimensionless
parameters: $k/K$ defines the stiffness of the interface,
and $H/a$ determines a ``capacity'' of the tectonic plate,
where the elastic stress is accumulated.

\section{Ageing of the contacts}
\label{app-ageing}

\begin{figure} 
\includegraphics[clip, width=8cm]{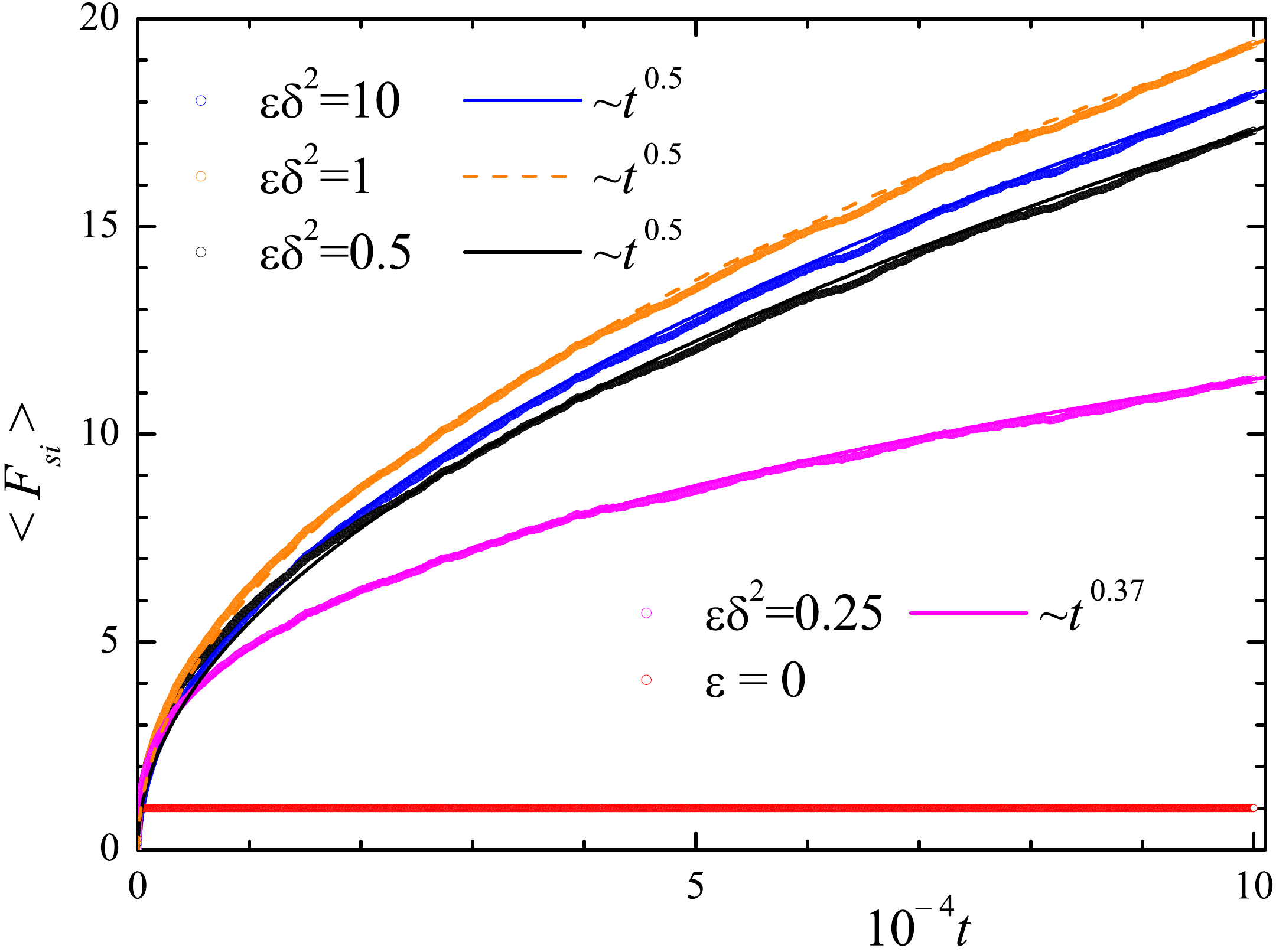} 
\caption{\label{A02}(color online):
Macro-contact ageing:
growth of the contact threshold $\langle F_{si} \rangle$
with its timelife for $\beta =0.3$, $\delta =0.1$ and
$\varepsilon =0$, 25, 50, 100 and 1000 (see legend).
Curves show power-law fits.
}
\end{figure}
The Langevin equations~(\ref{eq31}) for the thresholds $F_{si}$
have to be solved numerically.
For example, if at $t=0$ the threshold is zero,
then the average threshold value will grow with time
according to a power law $\langle F_{si} \rangle \propto t^{\nu}$
at short times as demonstrated in Fig.~\ref{A02},
where the value of the exponent $\nu \leq 0.5$ depends on the parameter $\varepsilon$.
%
For $\varepsilon \delta^2 < 0.5$
the average threshold
$\langle F_{si} \rangle = \int_0^{\infty} dF \, F {\cal P}_{c} (F;t)$
tends to saturate to a finite value at $t \to \infty$,
while for $\varepsilon \delta^2 \geq 0.5$ it keeps growing with time.

Now let us derive the corresponding Fokker-Planck equation
which will allow us to find some analytical results.
Equation~(\ref{eq31}) is equivalent to the stochastic equation~\cite{Gardiner}
\begin{equation}
\label{4.40a}
dF_{si}(t) = K(F_{si})\, dt + G \, dw \,,
\end{equation}
where
\begin{eqnarray}
\label{4.40bc}
\left\{
\begin{array}{ll}
\langle dw\rangle  = 0 \,,
\\
\langle dw(t)\, dw(t)\rangle  = dt \,.
\end{array}
\right.
\end{eqnarray}
Let us introduce the
distribution function ${\cal P}_c (F_{si},t|F_{si0},t_0)$
defined as the conditional probability that the contact $i$ has
the threshold $F_{si}$ at time $t$,
if at the previous time $t_0$ it had the value $F_{si0}$.
For an arbitrary function $u(F_{si})$,
its average value
$\langle u\rangle $ at time $t$ is equal to
\begin{equation}
\label{4.41}
\langle u \rangle =\int dq\, u(q)\, {\cal P}_c (q,t| \ldots ) \,,
\end{equation}
and its derivative over time is
\begin{equation}
\label{4.42}
\frac{d}{dt}\langle u\rangle =\int dq\, u(q)\,
\frac{\partial {\cal P}_c (q,t| \ldots )}{\partial t} \,.
\end{equation}

On the other hand, the differential of the function
$u(q)$ with an accuracy up to $dt$, with the help of
Eq.~(\ref{4.40a}) may be written in the form
\begin{eqnarray}
\label{4.43}
\begin{array}{ll}
du & = (\partial u/\partial F_{si})\, dF_{si} +
  \frac{1}{2}(\partial ^2u/
      \partial F_{si}^2)\, dF_{si}\, dF_{si}    \\
& = (\partial u/\partial F_{si})\,
      [K(F_{si})\, dt + G\, dw(t)] \\
& + \frac{1}{2}(\partial ^2u/
     \partial F_{si}^2)\, G^2 \, dw(t)\, dw(t) \,,
\end{array}
\end{eqnarray}
where the second derivative appears because
$\langle  dw\rangle \sim \sqrt{dt}$
in the stochastic equation.

Averaging Eq.~(\ref{4.43}) over time
using the Ito calculus~\cite{Gardiner},
dividing both sides of the equation by $dt$,
and taking into account Eqs.~(\ref{4.40bc}), we obtain
\begin{equation}
\label{4.44}
\frac{\langle du\rangle }{dt} =
\biggr\langle \frac{\partial u}{\partial F_{si}} K(F_{si}) \biggr\rangle  + \frac{1}{2}
\biggr\langle \frac{\partial ^2u}{\partial F_{si}^2}\, G^2 \biggr\rangle \,.
\end{equation}
The first term in the right-hand side of Eq.~(\ref{4.44})
may be rewritten as
\begin{eqnarray}
\label{4.45}
\biggr\langle \frac{\partial u}{\partial F_{si}}\, K(F_{si}) \biggr\rangle  =
\int dq\, {\cal P}_c (q,t| \ldots )\, K(q)\,
\frac{\partial u}{\partial q}
\nonumber \\
=-\int dq\, u(q)\, \frac{\partial }{\partial q}
[K(q)\, {\cal P}_c (q,t| \ldots )] \,,
\end{eqnarray}
where we also made the integration by parts.
In a similar way the second term in the right-hand side of
Eq.~(\ref{4.44}) may be transformed,
if we make the integration by parts two times.
Then, comparing the obtained expression for
$\langle du\rangle /dt$ with Eq.~(\ref{4.42})
and taking into account that the function $u(q)$
is an arbitrary one, we obtain that the distribution function
${\cal P}_c (F_{si},t| \ldots )$ must satisfy the following equation,
\begin{eqnarray}
\label{4.46}
\frac{\partial {\cal P}_c (F_{si},t| \ldots )}{\partial t} =
-\frac{\partial }{\partial F_{si}}
[K(F_{si})\, {\cal P}_c (F_{si},t|\cdots )]
\nonumber \\
+
\frac{1}{2}\frac{\partial ^2}{\partial F_{si}^2}
[G^2 {\cal P}_c (F_{si},t| \ldots )] \,,
\end{eqnarray}
which is the Fokker-Planck equation~(\ref{eq32}).


%
\end{document}